# High-Frequency Electron-Spin-Resonance Study of the Octanuclear Ferric Wheel CsFe$_8$

*Jan Dreiser,*[*,†] *Oliver Waldmann,*[*,†] *Graham Carver,*[‡] *Christopher Dobe,*[‡] *Hans-Ulrich Güdel,*[‡] *Høgni Weihe,*[¶] *and Anne-Laure Barra*[§]

[†]Physikalisches Institut, Universität Freiburg, D-79104 Freiburg, Germany, [‡]Department of Chemistry and Biochemistry, University of Bern, 3012 Bern, Switzerland, [¶]Department of Chemistry, University of Copenhagen, 2100 Copenhagen, Denmark, and [§]Grenoble High Magnetic Field Laboratory, CNRS, BP 166, 38042 Grenoble Cedex 9, France

*Authors to whom correspondence should be addressed. Email: jan.dreiser@physik.uni-freiburg.de; oliver.waldmann@physik.uni-freiburg.de

High-frequency ($f$ =190 GHz) electron paramagnetic resonance (EPR) at magnetic fields up to 12 T as well as Q-band ($f$ =34.1 GHz) EPR were performed on single crystals of the molecular wheel CsFe$_8$. In this molecule, eight Fe(III) ions, which are coupled by nearest-neighbor antiferromagnetic (AF) Heisenberg exchange interactions, form a nearly perfect ring. The angle-dependent EPR data allow for the accurate determination of the spin Hamiltonian parameters of the lowest spin multiplets with $S \leq 4$. Furthermore, the data can well be reproduced by a dimer model with a uniaxial anisotropy term, with only two free parameters $J$ and $D$. A fit to the dimer model yields $J = -15(2)$ cm$^{-1}$ and $D = -0.3940(8)$ cm$^{-1}$. A rhombic anisotropy term is found to be negligibly small, $E = 0.000(2)$ cm$^{-1}$. The results are in excellent agreement with previous inelastic neutron scattering (INS) and high-field torque measurements. They confirm that the CsFe$_8$ molecule is an excellent experimental model of an AF Heisenberg ring. These findings are also important within the scope of further investigations on this molecule such as the exploration of recently observed magnetoelastic instabilities.

**Introduction**

In recent years, antiferromagnetic (AF) molecular ferric wheels have attracted the interest of chemists and physicists alike because of their intriguing properties:[1] Thanks to their high symmetry, their magnetic energy-level structure and all deduced quantities are essentially equal to those of AF chains with periodic boundary conditions. However, the small size distinguishes them and was found to give rise to particular phenomena. In this context, it turned out that AF wheels are good physical model systems for the study of collective effects in small magnetic clusters. For instance, quantized spin waves or mesoscopic quantum coherence such as tunneling of the Néel vector have been observed in some of these wheels.[2] Very recently, the title compound [CsFe$_8${N(CH$_2$CH$_2$O)$_3$}$_8$]Cl, or CsFe$_8$ in short, gained additional interest because of the observation of magnetoelastic instabilities, which were attributed to a field-induced spin Jahn-Teller effect.[3] Furthermore, heterometallic AF wheels have been proposed and are currently experimentally investigated as qubit candidates.[4] Experimentally, AF ferric wheels have been studied using a variety of techniques, such as high-field torque magnetometry,[1g,1h] inelastic neutron scattering (INS),[1h] and nuclear magnetic resonance.[5] However, only a few electron paramagnetic resonance (EPR) investigations have been reported,[1f,4c,6] despite EPR and high-frequency EPR becoming standard in the field of molecular nanomagnets.[7]

Here, we report on multiple-frequency EPR experiments (Q-band and 190 GHz) on single crystals of the AF molecular wheel CsFe$_8$ revealing the full angular dependence of the EPR resonances. Figure 1 shows the molecule CsFe$_8$; it consists of eight Fe(III) ions ($s = 5/2$), which form an almost perfect ring with a crystallographic C$_4$ symmetry axis ($z$ axis) perpendicular to the wheel plane. The CsFe$_8$ wheel has been carefully studied by high-field torque magnetometry and INS.[2h,8] The collective result of these experiments was that the CsFe$_8$ molecule can be very well described by a generic spin Hamiltonian with dominant nearest-neighbor AF Heisenberg couplings plus uniaxial single-ion anisotropy terms. However, the INS technique provides about ten times less experimental resolution as compared to EPR. Furthermore, because of the particular spin structure realized in AF wheels, the INS transitions within a spin multiplet have much weaker intensities than the transitions between different multiplets (although both are allowed by the INS selection rules).[8b] Both aspects strongly motivated the present EPR experiments, which aimed at obtaining a more precise and complete knowledge of the energy level structure. The results will show that the most simple, generic spin Hamiltonian describes $CsFe_8$ with high accuracy, which implies that magnetismwise $CsFe_8$ is highly symmetric. This insight should be of great relevance to better understand, e.g., the spin Jahn-Teller effect in this molecule. In addition to the extraction and precise determination of spin Hamiltonian parameters, the data allow for the clear observation of $S$ mixing.[9]

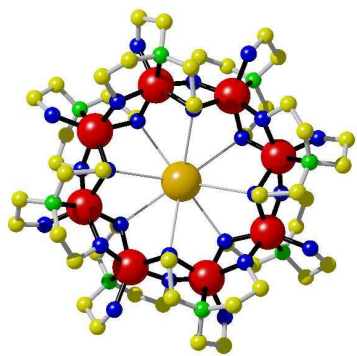

**Figure 1.** View of the molecular structure of the $CsFe_8$ molecule. Fe(III) ions are drawn as red spheres. H atoms were omitted.

**Experimental Techniques**

**Sample Synthesis and Collection of EPR Data.** A powder sample of $CsFe_8$ was prepared as previously described,[1d] and subsequent recrystallization of the powder from a 1:1 mixture of $CHCl_3$ and $CH_2Cl_2$ by vapor diffusion of pentane yielded amber-colored platelets. High-frequency EPR ($f$ = 190 GHz) measurements on a large (~5×4×1 $mm^3$) single crystal of $CsFe_8$ were performed at the Grenoble

High Magnetic Field Laboratory at the CNRS in Grenoble, France. Magnetic fields of up to 12 T were applied at temperatures of 5 and 15 K, respectively; the measurement apparatus has been described in detail previously.[7b] Q-band EPR spectra ($f$ = 34.147 GHz) of a CsFe$_8$ single crystal (~2×1×0.1 mm$^3$) were measured using a Bruker Elexsys E500 spectrometer equipped with an Oxford Instruments cryostat.

**Simulation and Fitting of EPR Data.** Relevant energy levels were obtained with in-house-written software, which used exact numerical diagonalization techniques (both full diagonalization and sparse-matrix techniques).[10] The EPR simulations and fits were performed using the programs *sim* and *esrfit*.[11] For fitting the function $\chi^2 = \sum_{i=1}^{N_{obs}} (B_{obs,i} - B_{calc,i})^2 / \sigma_i^2$ was minimized, with $B_{obs,i}$ and $B_{calc,i}$ the $i$th observed and calculated resonance magnetic field, respectively. The denominator $\sigma_i$ represents the uncertainty on the $i$th resonance magnetic field; we chose 300 G, i.e., a typical bandwidth in the 190 GHz spectra, for all observations.

**Theoretical Modeling and Spin Hamiltonians**

Our modeling of the EPR data is based on various (effective) spin Hamiltonians, which operate at different levels of sophistication. The most fundamental spin Hamiltonian, which we will call the *microscopic spin Hamiltonian*, describes the CsFe$_8$ wheel at the level of the individual spin centers and their interactions. The low-lying energy states relevant to the EPR experiment can also be handled by a less fundamental effective Hamiltonian, which we will call the *sublattice Hamiltonian*. It grasps the consequences of the AF correlations in the low-lying states correctly and with high accuracy, yet simplifies the treatment considerably. Although the strong-exchange limit is not well realized for the lowest-lying states (because of $S$ mixing, *vide infra*), it happens that the states may, nevertheless, be grouped by their total spin quantum number $S$ into spin multiplets, which is true, in particular, for the higher excited states. As a result the widely used approach of introducing a spin Hamiltonian for each spin multiplet, which we will call the *spin-multiplet Hamiltonian*, is employed. This approach is essentially phenomenological and represents the lowest level of approximation.

Because of the dominant AF coupling in CsFe$_8$ the ground state is nonmagnetic and belongs to $S = 0$. Higher states become populated depending on the temperature of the experiment; in the present work the lowest multiplets with $S = 1 - 4$ are observed. Because of the EPR selection rule $\Delta S = 0$, only the transitions within the spin multiplets are allowed. In the following sections, the three Hamiltonians will be described in more detail.

**Microscopic Spin Hamiltonian.** As indicated in the Introduction, the previous experiments on CsFe$_8$ could be well described by an AF Heisenberg spin ring with a small uniaxial anisotropy along the wheel axis. Together with a Zeeman term, the spin Hamiltonian reads

$$\hat{H} = -J\left[\sum_{i=1}^{N-1}\hat{\mathbf{s}}_i \cdot \hat{\mathbf{s}}_{i+1} + \hat{\mathbf{s}}_N \cdot \hat{\mathbf{s}}_1\right] + D\sum_{i=1}^{N}\left[\hat{s}_{i,z}^2 - \frac{1}{3}s_i(s_i+1)\right] + g\mu_B\sum_{i=1}^{N}\hat{\mathbf{s}}_i \cdot \mathbf{B} + \hat{H}', \qquad (1)$$

with $J < 0$ the magnetic coupling strength, $N = 8$ the number of magnetic sites, $\hat{s}_i$ the spin operator of the $i$th ion with spin $s = 5/2$, $D$ the single-ion anisotropy oriented along the wheel's $z$ axis, and $g$ the isotropic single-ion $g$ factors. The term $\hat{H}'$ captures all other possible terms such as biaxial anisotropy ($E$ terms), higher-order interactions ($B^0_4$ or $B^4_4$ terms, anisotropic exchange, Dzyaloshinsky-Moriya interactions), and so on. It is one of the goals of this work to get an estimate for the importance of such terms in CsFe$_8$. Of particular interest is the $E$ term, as it would be the leading contribution to $\hat{H}'$ if the symmetry of the molecule would be lower than the crystallographic C$_4$ symmetry. The experimental findings will demonstrate that the additional terms lead only to small corrections, i.e., that the major physics is determined by the $J$ and $D$ terms.

The microscopic spin Hamiltonian is the basis for the description of small spin clusters such as the CsFe$_8$ wheel. It often allows one to accurately reproduce experimental data; however, the physics that underlies the structure of the resulting energy states and wave functions is usually less obvious. For the even-numbered AF Heisenberg spin rings, considerable theoretical and experimental work in recent years has elucidated the main physical aspects:[2c-2e] The low-energy spectrum (for small numbers of sites $N$) is very well represented by the rotational-band picture, according to which the lowest-lying states of each spin $S$ sector form the so-called $L$ band, followed by the $E$ band corresponding to (quantized) spin-wave states, and the quasi continuum.[2d] It is sketched in Figure 2a for the limiting case of a pure Heisenberg interaction, i.e., $D = 0$ and $B = 0$.

**Sublattice Hamiltonian.** An effective Hamiltonian is obtained by reducing the two AF sublattices in the wheel to a dimer of spins $S_A$ and $S_B$, each of length $S_{A,B} = Ns/2 = 10$ with an effective AF coupling $j$ and uniaxial anisotropy $d$. The Hamiltonian reads

$$\hat{H}_{AB} = -j\,\hat{\mathbf{S}}_A \cdot \hat{\mathbf{S}}_B + d\left[\left(\hat{S}_{A,z}^2 + \hat{S}_{B,z}^2\right) - \frac{1}{3}\left[S_A(S_A+1) + S_B(S_B+1)\right]\right] + g\mu_B\left(\hat{\mathbf{S}}_A + \hat{\mathbf{S}}_B\right)\cdot \mathbf{B}\,, \qquad (2)$$

with the relations $j = 0.5536\,J$ and $d = 0.1870\,D$, which were determined such that the exact low-energy spectra of eqs 1 and 2 match.[12] This Hamiltonian describes the states of the $L$ band but does not capture the states of the $E$ band or quasi continuum. However, these missing states lie above ca. 60 cm$^{-1}$ and thus only become important at high temperatures. For the $L$-band states, the values of the energies and matrix elements are excellently reproduced by eq 2, even in the presence of large magnetic anisotropy.[12] For the states relevant in this work, this is shown in Figures 2c and 2d, which compare the exact energies of eqs 1 and 2. Hence, the sublattice Hamiltonian provides an adequate and highly accurate tool for analyzing the EPR data and capturing the relevant effects, including $S$ mixing. Moreover, it is technically advantageous because of the much smaller dimension of the Hilbert space (441), which enables quick simulations and allows full least-squares fits to the EPR data. It should be noted that in this approach no additional parameters are introduced compared to the microscopic spin Hamiltonian.

**Spin-Multiplet Hamiltonian.** In general, the EPR spectrum of an exchange-coupled cluster in the strong-exchange limit can be described as the superposition of the EPR spectra of the individual spin multiplets resulting from the exchange coupling.[13] The anisotropy terms in the microscopic spin Hamiltonian, such as single-ion anisotropy, then lead to anisotropy terms in the effective Hamiltonian for each individual spin multiplet, i.e., the spin-multiplet Hamiltonians, as sketched in Figure 2b. A further consequence of this approach is the appearance of higher-order terms in the spin-multiplet Hamiltonian (for multiplets with $S > 1$) due to $S$ mixing.[9] The Hamiltonian for an individual spin multiplet with spin $S$ reads

$$\hat{H}_S = \Delta_S + D_S\left[\hat{S}_z^2 - \frac{1}{3}S(S+1)\right] + B_{4,S}^0 \hat{O}_4^0 + g_S \mu_B \hat{\mathbf{S}} \cdot \mathbf{B}, \tag{3}$$

where $\Delta_S$ is the center-of-gravity energy of the multiplet, $D_S$ the uniaxial anisotropy, $B_{4,S}^0$ a fourth-order anisotropy term, and $g_s = g$ the isotropic g factor. In principle, further (higher-order) terms may appear; eq 3 has been restricted to the terms relevant in this work. A connection of eq 3 to a higher-level Hamiltonian may be made by fitting to simulated data.[14] The center-of-gravity $\Delta_S$ does not affect the EPR resonance fields and, hence, cannot directly be determined from an EPR spectrum but may indirectly be inferred from the temperature dependence of the intensity of the EPR transitions. The present experiments did not provide accurate enough temperature dependencies; the parameter $\Delta_S$ was thus irrelevant for our analysis.

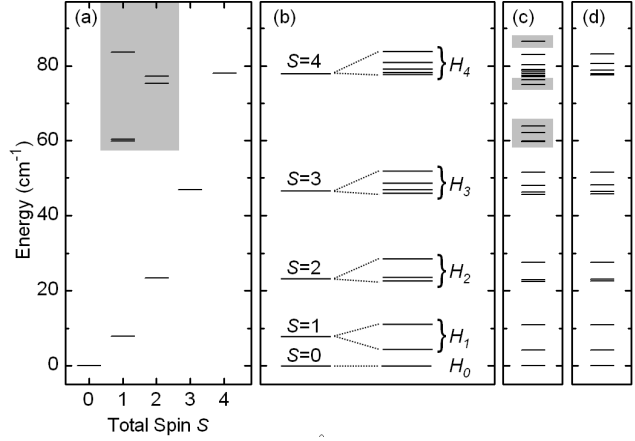

**Figure 2.** Calculated energy spectrum for CsFe$_8$ for different approximations ($J$ = -14.6 cm$^{-1}$, $D$ = -0.394 cm$^{-1}$). (a) Energies vs total spin $S$ of the microscopic spin Hamiltonian without anisotropy, $D$ = 0. The lowest states for each $S$ belong to the $L$ band, the states in the shaded region belong to the $E$ band. (b) Splitting of the spin multiplets of the $L$ band by the uniaxial anisotropy $D$ and the association with spin-multiplet Hamiltonians. (c) Calculated spectrum from the microscopic spin Hamiltonian using eq 1, including anisotropy. The shaded regions again indicate states of the $E$ band. (d) Calculated spectrum from the sublattice Hamiltonian using eq 2.

**Experimental Results and Analysis**

Figure 3 shows the recorded EPR spectra for various angles $\beta$ between the applied magnetic field and molecular $z$ axis. The upper panel presents the high-frequency experimental data; the lower panel shows the Q-band data. For both frequencies, a large number of transitions were observed, which shift significantly with the angle (in Figure 3, only the strongest resonances may be identified). For each data curve, the field positions of the resonances were extracted, as indicated for one example curve in Figure 4. The obtained resonance fields are plotted in Figures 5 and 6 as a function of angle $\beta$ for the high-frequency and Q-band experiments, respectively.

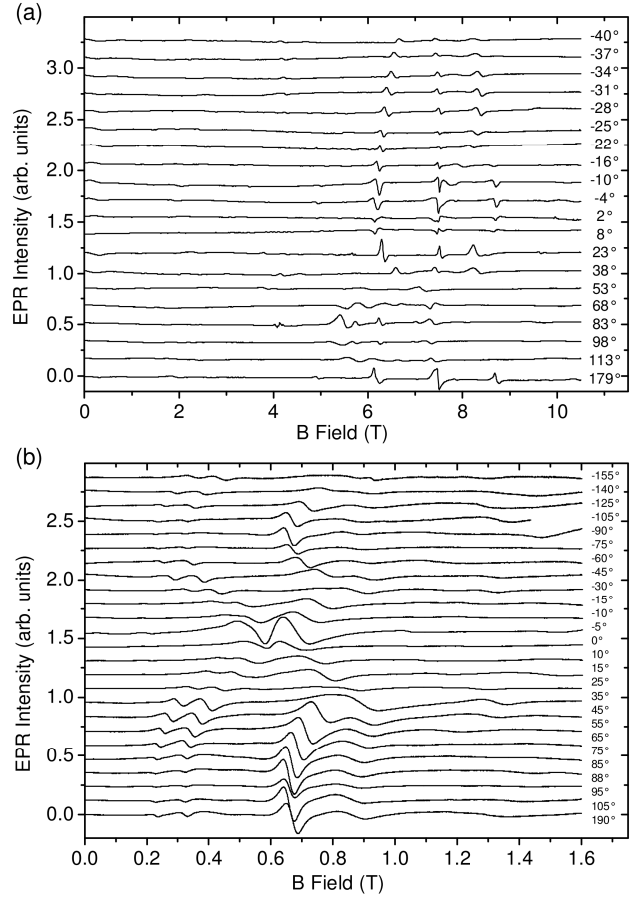

**Figure 3.** Measured EPR spectra for the angles $\beta$ indicated in the panels (with varying angle increment) taken at a frequency of (a) $f = 190$ GHz and (b) $f = 34.1$ GHz. The temperatures were $T = 15$ K in (a) and $T = 25$ K in (b), respectively.

The extracted resonance positions were then used in the analysis, which consisted of two steps. First, the parameter space was scanned in a reasonable range in order to find first guesses. It in fact required extensive numerical calculations in order to find a good parameter set at all, as the EPR spectrum reacted sensitively on their values. Only one spot of good parameters was found. Then, in the next step, the experimental resonance positions were least-squares fit to the calculated positions, using the previously found first guesses as starting values. In a fit all the resonances for one EPR frequency were used simultaneously, however, high-frequency and Q-band data were analyzed separately. This not only simplified the fitting procedure, but by comparing the resulting best-fit parameters provides a check of the modeling. For each frequency the resonance fields were fitted using both the spin-multiplet Hamiltonian approach (with an appropriate number of spin multiplets) and the sublattice Hamiltonian. Although in principle feasible, a least-squares fit to the EPR data was not attempted with eq 1, because the additional complexity due to the applied magnetic field would have made the calculations time consuming, and because the sublattice Hamiltonian provides a highly accurate yet much faster approach.

In the fits the $g$ factor was fixed to $g = 2.0$. Slightly smaller $g$ values were tested (the smallest we considered was $g = 1.95$), but did not improve the fits within statistical significance. The procedure yielded four best fits, which are presented in Figures 5 and 6.

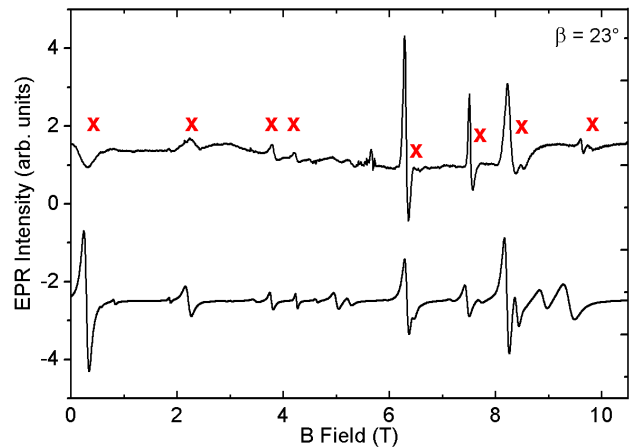

**Figure 4.** High-frequency EPR spectrum for $\beta = 23°$ at a temperature of $T = 15$ K. The top curve shows the experimental spectrum, where the crosses indicate the resonances used in the data analysis. The bottom curve shows the simulated spectrum as calculated from eq 2 using the parameters given in the text.

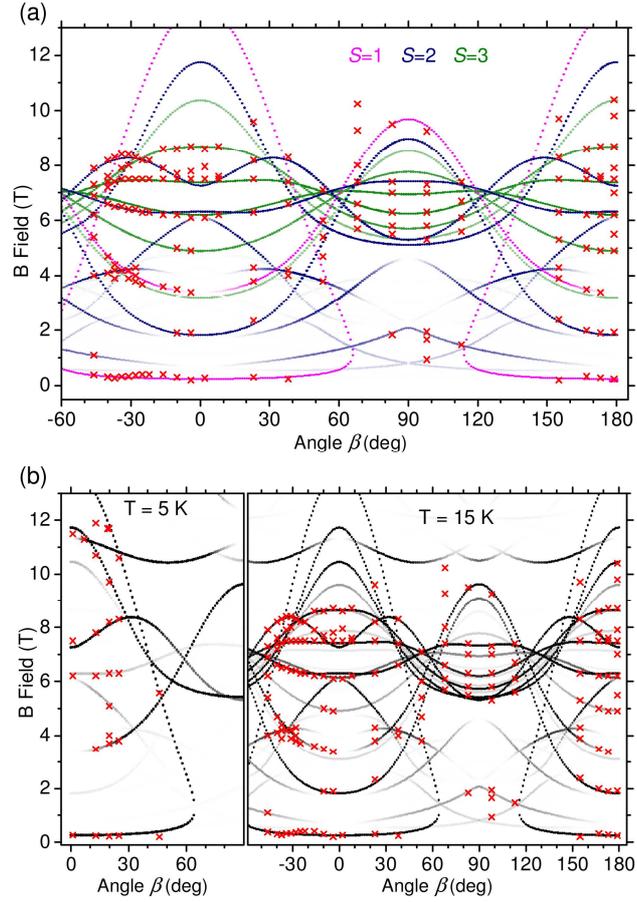

**Figure 5.** Experimental and simulated EPR resonance fields at a frequency of $f$ = 190 GHz as a function of angle $\beta$ (temperatures as indicated). In both panels, the red crosses mark the experimentally observed resonance fields. The dotted curves represent the fit results using (a) the spin-multiplet Hamiltonians eq 3 for each of the three lowest multiplets with $S = 1 - 3$, and (b) the sublattice Hamiltonian eq 2. The obtained best-fit parameters are given in the text. The calculated intensity of the transition is encoded in the color intensity of the dotted curves (light = weak; dark = strong).

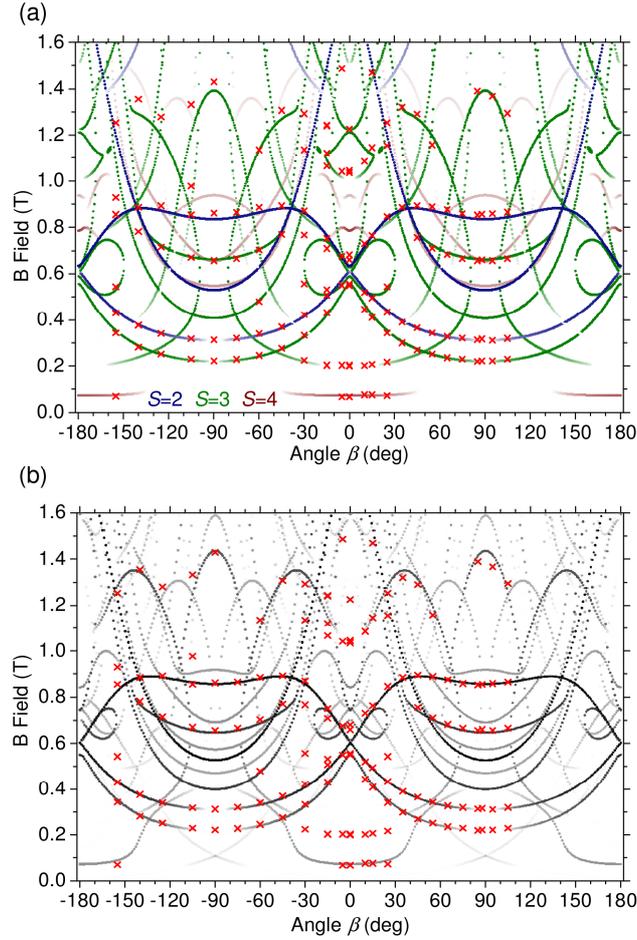

**Figure 6.** Experimental and simulated EPR resonance fields at a frequency of $f = 34.1$ GHz as a function of angle $\beta$ (temperature was $T = 25$ K). In both panels, the red crosses represent the experimentally observed resonance fields. The dotted curves represent the fit results using (a) the spin-multiplet Hamiltonian eq 3 for each of the three multiplets with $S = 2 - 4$, and (b) the sublattice Hamiltonian eq 2. The obtained best-fit parameters are given in the text. The calculated intensity of the transition is encoded in the color intensity of the dotted curves (light = weak; dark = strong).

The resonances in the high-frequency data are essentially due to the three lowest multiplets with $S = 1 - 3$. In the Q-band experiment, the transitions within the $S = 1$ multiplet could not be observed because a frequency of 34.1 GHz is too small. However, the transitions within the $S = 4$ multiplet were detected because of the larger sensitivity in this experiment. Since the $S = 4$ multiplet is located at an energy similar to that of the lowest *E* band states (Figure 2c), sample calculations with the microscopic Hamiltonian were performed in order to check for transitions within the *E* band, but as in the previous EPR experiments on AF ferric wheels,[6] such transitions could not be identified. As reported for the AF wheel NaFe$_6$,[6c] the lifetime of the *E*-band states is probably too short and the related EPR resonances too broad for these transitions to be seen in EPR experiments. It is also noted that transitions between spin multiplets were not observed, although because of *S* mixing, they may, in principle, gain EPR intensity.[9e] However, numerical simulations revealed that in CsFe$_8$ their intensity is too small to be detected in our experiments. The assignment of the observed resonances to intramultiplet transitions is shown for one high-frequency spectrum in Figure 7. Also, the simulated spectrum is presented for comparison. The resonance fields of all observed transitions are reproduced excellently; the intensities are reproduced moderately because of line broadening and instrumental effects.

For the parameters of the spin-multiplet Hamiltonians the following best-fit values were obtained:

190 GHz:  $D_1 = 6.558(3)$ cm$^{-1}$,

$D_2 = 1.39(2)$ cm$^{-1}$,     $B^0_{4,2} = 7.9(7) \times 10^{-3}$ cm$^{-1}$,

$D_3 = 0.64(1)$ cm$^{-1}$,     $B^0_{4,3} = 2.9(8) \times 10^{-4}$ cm$^{-1}$,

34.1 GHz:  $D_2 = 1.475(4)$ cm$^{-1}$,  $B^0_{4,2} = 8.0(2) \times 10^{-3}$ cm$^{-1}$,

$D_3 = 0.501(6)$ cm$^{-1}$,  $B^0_{4,3} = -3.9(6) \times 10^{-4}$ cm$^{-1}$,

$D_4 = 0.403(4)$ cm$^{-1}$,  $B^0_{4,4} = 0.0(1) \times 10^{-4}$ cm$^{-1}$.

The agreement of the parameters for 190 GHz and Q band is very good for the $S = 2$ multiplet, but for the $S = 3$ multiplet the $D_3$ parameters differ quite significantly, and the $B^0_{4,3}$ parameters are even of opposite sign. This will be explained below. Of interest is also the large value of the $B^0_{4,2}$ parameter of the $S = 2$ multiplet. For the sublattice Hamiltonian, the best-fit parameters were

190 GHz:   $J = -15(2)$ cm$^{-1}$, $D = -0.394(1)$ cm$^{-1}$,

34.1 GHz:   $J = -17(2)$ cm$^{-1}$, $D = -0.413(7)$ cm$^{-1}$,

which are in very good agreement with each other and with the previous INS results $J = -14.6(2)$ cm$^{-1}$ and $D = -0.405(8)$ cm$^{-1}$.[2h,8b,8c] We performed also additional fits that included a biaxial anisotropy term $E\left(\left(\hat{S}_{A,x}^2 - \hat{S}_{A,y}^2\right) + \left(\hat{S}_{B,x}^2 - \hat{S}_{B,y}^2\right)\right)$ in the sublattice Hamiltonian. This yielded $E = 0.000(2)$ cm$^{-1}$; i.e., within experimental resolution, a biaxial anisotropy is zero.

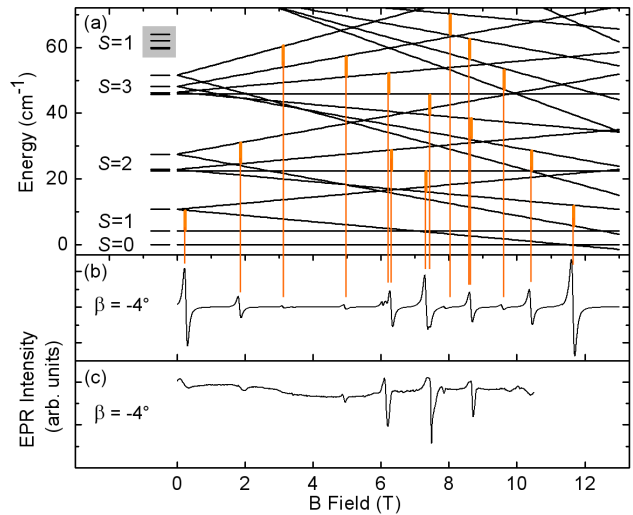

**Figure 7.** (a) Energy level scheme as a function of an applied magnetic field along the molecular $z$ axis (solid black lines). The scheme was calculated using the sublattice Hamiltonian eq 2 with $J = -14.6$ cm$^{-1}$, $D = -0.394$ cm$^{-1}$, and $g = 2.0$. The resonance positions for the high-frequency measurements at $f = 190$ GHz are indicated by vertical lines. The horizontal dashes on the left indicate the zero-field low-energy spectrum of the microscopic spin Hamiltonian eq 1. Also, their association to spin multiplets is given. The shaded region marks states belonging to the $E$ band. (b) Simulated EPR spectrum at $T = 15$ K for a field almost parallel to the molecular $z$ axis as calculated with the sublattice Hamiltonian eq 2 and parameters identical with those in panel a. The thin vertical lines are guides to the eye relating the simulated resonance positions to the energy-level scheme in panel a. (c) Experimental high-frequency EPR spectrum at $T = 15$ K for a field almost parallel to the molecular $z$ axis.

**Discussion**

In general, the fits presented above reproduce the experimental data excellently. However, a very careful inspection (e.g., Figures 5 and 6) reveals that the 190 GHz data are significantly better reproduced than the Q-band data. In addition, the spin-multiplet Hamiltonian parameters showed significant differences, in particular, for the $S = 3$ multiplet. Both findings may be explained by the largely different EPR frequencies, which determine the energy scale that is probed in the experiments. In the Q-band experiment, the frequency and applied magnetic fields are much smaller, which implies a smaller typical energy scale, while in the 190 GHz experiment, for most of the resonances the energy level splitting is dominated by the Zeeman energy. Hence, in the Q-band experiment the EPR resonance positions should be much more sensitive to small shifts in the energies because of terms not yet accounted for in the spin Hamiltonian, i.e., the terms in $\hat{H}'$ in eq 1 (and corresponding terms in eq 2 or eq 3). The above findings thus imply that additional terms besides the $J$ and $D$ terms are present in CsFe$_8$, but are so weak that the Q-band resonances are slightly and the 190 GHz resonances essentially not affected. The discrepancy between the $S = 3$ spin-multiplet parameters might serve as a quantitative estimate for the strength of these terms. The argument is strengthened by the fits with the sublattice Hamiltonian, which provides equally good fits as those with the spin-multiplet Hamiltonian but involves only two free parameters. Hence, a crucial term cannot be missed in the description. Thus, our data show that additional terms in the Hamiltonian for CsFe$_8$ are present but very weak; i.e., they are significant enough to somewhat affect the Q-band resonances but are too weak to be clearly discerned in our Q-band experiment and even less so in the 190 GHz experiment.

As mentioned before, the possibility of a biaxial anisotropy term was explicitly considered. Such a term would be forbidden for the crystallographic symmetry of the CsFe$_8$ molecule. However, the actual molecular symmetry may be lower because of various effects (disorder, imperfections, strains, etc.), giving rise to an $E$ term. Prominent examples in this context are the Mn$_{12}$acetate single-molecule magnet or the $Cr_8$ AF wheel.[15] Testing for an $E$ term, hence, provides a critical check on the molecular symmetry. For $CsFe_8$, the previous INS measurements yielded an upper bound of $E/D < 0.1$;[8c] the present EPR study dramatically lowers it to $E/D < 5\times10^{-4}$. Apparently, in $CsFe_8$, deviations from $C_4$ symmetry are if at all very small. This finding will allow a better understanding of the field-induced spin Jahn-Teller effect.[3] Here, the molecule undergoes a spontaneous distortion, lowering its symmetry, at very low temperatures ($T < 1$ K) for magnetic fields close to a level crossing. Clearly, the structure of the molecule is a key ingredient in describing this phenomenon; the present results shed some important light on the symmetry aspects of the problem.

In the energy-level diagrams shown in Figures 2 and 7, the two lowest excited levels are grouped into a $S = 1$ multiplet, although its zero-field splitting (ZFS) of about 6.6 cm$^{-1}$ is so large that the multiplet structure is not obvious from the energy spectrum. In fact, the first excited level lies closer to the ground state than to the second excited level. As discussed already in refs 8b and 8c, this signals strong $S$ mixing[9] or the breakdown of the strong-exchange limit, respectively (which in a technically strict sense renders the notion of a spin multiplet meaningless). $S$ mixing is intrinsically included in the higher-level spin Hamiltonians but not in the spin-multiplet Hamiltonian; there it needs to be accounted for by adding additional higher-order anisotropy terms.

In the case of $CsFe_8$, the strong $S$ mixing is evident from the very large ZFS of the $S = 1$ multiplet. However, its effects can also be seen in the higher-lying spin multiplets, in particular the $S = 2$ multiplet. As a matter of fact, $S$ mixing is strongest for the lowest states but becomes weaker with increasing energy, as demonstrated by the clearer grouping of the states into multiplets in Figure 7. For instance, $S$ mixing drastically affects the sublevel structure of the $S = 2$ multiplet and leads to the situation that the $M_S = \pm1$ states lie at almost the same energy as the $M_S = 0$ state (22.9 and 22.4 cm$^{-1}$ compared to 27.6 cm$^{-1}$ of the $M_S = \pm2$ states), which gives rise to the large $B^0_{4,2}$ value. If one fits the energies of the $S = 2$ multiplet as calculated with the sublattice (or microscopic) Hamiltonian to the spin-multiplet Hamiltonian, one obtains $D_2 = 1.397$ cm$^{-1}$ and $B^0_{4,2} = 7.66\times10^{-3}$ cm$^{-1}$, which are remarkably close to the experimentally determined values. This explicitly demonstrates that $S$ mixing is fully accounted for by the sublattice Hamiltonian and, moreover, that the large experimentally observed $B^0_{4,2}$ value is indeed due to $S$ mixing.

Furthermore, it is stressed again that the analysis is based on the EPR resonance fields and, hence, is insensitive to the center-of-gravity energies $\Delta_S$ of the multiplets. These are governed by the exchange couplings. Therefore, the analysis should not reveal information on the coupling constant $J$ except through the indirect relative shifts of intramultiplet levels by the $S$-mixing mechanism.[9c] The good agreement of the $J$ values determined from the fits of the sublattice Hamiltonian to the 190 GHz and Q-band EPR data and the excellent agreement with the $J$ value inferred from INS clearly demonstrates the strong consistency of the modeling. The comparison between EPR and INS is, in particular, compelling because in the EPR experiment only energy splittings *within* spin multiplets were probed, while in the INS experiment only splittings *between* spin multiplets were observed.[8b,8c]

It is actually one of the more striking aspects of the INS spectrum in AF wheels that the *intra*multiplet transitions were not observed, although allowed by the INS selection rules. However, their INS intensities are orders of magnitude weaker than those of *inter*multiplet transitions (as revealed by numerical simulations based on eqs 1 and 2 in ref 8c). As argued in ref 8b, this is a consequence of a sublattice structure with two mesoscopically sized spins on each sublattice ($S_A = S_B = 10$), which greatly enhances the *inter*multiplet transitions rendering the *intra*multiplet transitions too weak to be observed in the INS experiments. The transition between the first and second excited levels is of particular interest in this context because it should easily have been seen in the INS experiments. Figure 8 shows a representative high-frequency EPR spectrum and, for better comparison, depicts again the three relevant levels as well as reproduces relevant INS data. The *inter*multiplet transitions are denoted as I and II; the *intra*multiplet transition is denoted as α. In the INS data, transitions I and II are clearly observed, while transition α appears to be absent. However, in the EPR spectrum, transition α is clearly detected, unequivocally confirming its existence. Hence, the combined EPR and INS data provide direct experimental proof of a hallmark feature of AF wheels (and similar systems), namely, the mesoscopic AF sublattice structure.[2g,8b]

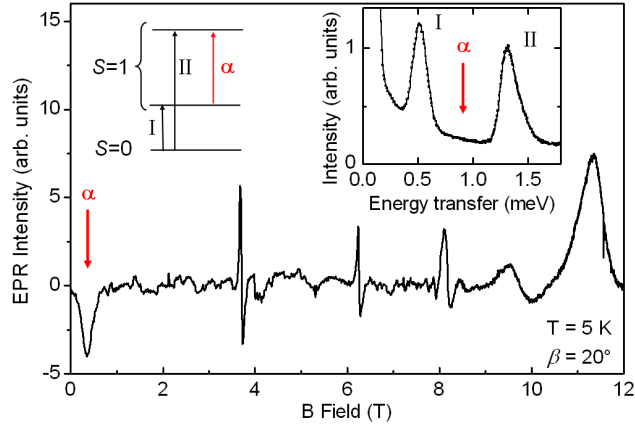

**Figure 8.** High-frequency EPR spectrum taken at $T = 5$ K and $\beta = 20°$. The upper left inset sketches the lowest three energy levels in CsFe$_8$, the arrows indicate the intramultiplet transition $\alpha$, and the intermultiplet transitions I and II. The upper right inset shows an INS spectrum taken at $T = 9.7$ K. The transition $\alpha$ is too weak to be detected in the INS experiment, but clearly present in the EPR data.

**Conclusion**

In conclusion, high-frequency (f = 190 GHz) and Q-band (f = 34.1 GHz) EPR data on single crystals of CsFe$_8$ were recorded as a function of the orientation of the applied magnetic field. The data were fitted using the spin Hamiltonian approach at different levels of sophistication, specifically the sublattice and the spin-multiplet Hamiltonian, which both yielded excellent fitting results. The comparison of the best-fit parameters for the 190 GHz and Q-band data revealed small differences, which are attributed to additional terms in the Hamiltonian that have so far been neglected. However, the experimental data and comparison to the previous INS experiments demonstrate that these additional terms are small. The possibility of a biaxial anisotropy was explicitly considered and the *E/D* ratio was found to be very small, $E/D < 5 \times 10^{-4}$. The results demonstrate that the CsFe$_8$ molecule constitutes a rare example of a nearly perfect ring from the magnetic topology point of view. Furthermore, the presence of *S* mixing was clearly evidenced by the need of large fourth-order terms in the spin-multiplet Hamiltonian description, though such terms are absent in the sublattice or microscopic spin Hamiltonian.

**Acknowledgement**

This work was funded by the Deutsche Forschungsgemeinschaft. Access to the GHMFL through the European Commission Program ("Transnational Access - Specific Support Action" Program - Contract n° RITA-CT-2003-505474) is acknowledged.